\documentclass[letterpaper]{article} % DO NOT CHANGE THIS
\usepackage{aaai2026}  % DO NOT CHANGE THIS
\usepackage{times}  % DO NOT CHANGE THIS
\usepackage{helvet}  % DO NOT CHANGE THIS
\usepackage{courier}  % DO NOT CHANGE THIS
\usepackage[hyphens]{url}  % DO NOT CHANGE THIS
\usepackage{graphicx} % DO NOT CHANGE THIS
\usepackage{natbib}  % DO NOT CHANGE THIS AND DO NOT ADD ANY OPTIONS TO IT
\usepackage{caption} % DO NOT CHANGE THIS AND DO NOT ADD ANY OPTIONS TO IT
\usepackage{algorithm}
\usepackage{algorithmic}

\usepackage{newfloat}
\usepackage{listings}
\DeclareCaptionStyle{ruled}{labelfont=normalfont,labelsep=colon,strut=off} % DO NOT CHANGE THIS
\floatstyle{ruled}
\newfloat{listing}{tb}{lst}{}
\floatname{listing}{Listing}
\title{From Forensics to Ecosystems: \\ Rethinking Watermarks for Generative AI Oversight}
\author {
    Daniel Susser\textsuperscript{\rm 1}\thanks{Authors listed alphabetically. All contributed equally. The authors thank Calvin Qiu for invaluable research assistance.},
    John Thickstun\textsuperscript{\rm 2},
    Gili Vidan\textsuperscript{\rm 1}
}
\affiliations {
    \textsuperscript{\rm 1}Dept. of Information Science, Cornell University\\
    \textsuperscript{\rm 2}Dept. of Computer Science, Cornell University\\
    \{susser, jthickstun, gvidan\}@cornell.edu
}

\usepackage{bibentry}
\begin{document}

\maketitle

\begin{abstract}
The arrival of generative AI as a cheap, widely accessible commercial service, and the tidal wave of AI-generated synthetic content it has unleashed, have provoked deep epistemic and social anxieties and raised difficult governance questions that policymakers are struggling to address. One approach that has attracted both enthusiasm from regulators and skepticism from researchers is digital watermarking. Signals embedded in a synthetically-generated piece of content indicating that it was AI-generated---possibly even identifying the specific systems that generated it---appear to offer a path toward mitigating risks of genAI that avoids the downsides of more interventionist strategies. But critics warn that watermarks may prove technically brittle, epistemically ambiguous, and politically ineffectual tools. In this paper, we explore the challenges and opportunities of using digital watermarking for AI governance, paying special attention to the specific problem of watermarking AI-generated text. We argue that such critiques often treat the problem of identifying synthetic content as an isolated forensic question. Instead, we propose reconceptualizing digital watermarks as tools for understanding the impacts of synthetic content on media ecosystems, rather than reliably identifying individual pieces of synthetic content. Such an ``ecosystems approach'' more effectively utilizes the features of watermarks. And while this approach raises its own governance challenges, we argue that they are more tractable than the challenges of using watermarks for digital forensics.
\end{abstract}

\section{Introduction}

In 2018, tech journalist Max Read asked, ``how much of the internet is fake?'' His short piece on the eerie sense that a lot of online content is not only being generated by non-humans, but is also intended for consumption by non-humans, answered pithily, ``turns out, a lot of it, actually.'' Read described a hypothetical inflection point called ``the inversion'' at which the non-human traffic and content would become so pervasive, detection tools would start treating it as the authentic metric against which human action would be measured (and censored) \cite{read2018internet}. In the years since, the rise of commercially available generative AI (genAI) made the hypothetical seem imminent, shaping much of the popular media discussion around AI-generated or ``synthetic''\footnote{In both scholarly and popular news accounts, there is often slippage and contested use of various terms for content generated by AI tools, from ``deepfakes'' to ``generative AI content'' (and indeed many forms of digitally editing and manipulating media or text that predate the current focus on large language models). Throughout the paper, we choose to use ``synthetic content'' as an intentionally loose term, described by the US National Institute for Standards and Technology (NIST) as information ``significantly altered or generated by algorithms, including AI.'' \cite{NISTAI100-4}
} content, described by philosopher Joshua Habgood-Coote as the ``epistemic apocalypse narrative'' \cite{Habgood-Coote2023-HABDAT-2}. The detection conversation also shifted, from monitoring online traffic patterns, to examining the mechanism for content production itself. If only we could confidently know which pieces of media were produced by AI, perhaps we could also confidently know what to do about them.

Focusing on the means of content production sidesteps a variety of thorny, value-laden questions of content moderation, institutional trust, and governance structures in favor of what at first appears as a narrower, technical, and hopefully solvable question---was this content generated using AI tools?\footnote{Critiques of the ``epistemic apocalypse narrative'' have pointed out that a focus on which media was produced by AI (and therefore is often categorized as ``fake'') eschews questions about which uses of genAI or more broadly fakery are legitimate, desirable, or harmful, ignores the long history of media manipulation, and presupposes the utility of the label. See, for example, \citet{immerwahr2023doomsayers} in his review of \citet{ScheirerWalter2024Ahof}.} But, as we discuss below, even this framing of the question around the production of media still raises deep epistemic and normative questions around governance and is anything but technically straightforward. One of the most popular proposals for such a technical solution is the use of digital watermarking to tag and therefore allow for detection and disclosure of synthetic content. A digital watermark is a way to embed additional information in digital media (from videos to text) to indicate its provenance. On a technical level, audio, video, and image watermarks function differently from watermarks embedded in textual output, as we elaborate below, but we discuss the possibility of watermarking as an umbrella term across these cases in order to surface common uses and limitations.

While representing only one approach to ``digital content transparency,'' digital watermarking has been proposed by both policymakers and by commercial AI model developers and providers.\footnote{It is worth distinguishing between model developers and model providers, which may not always be the same entity. For example, OpenAI is both a model developer and model provider, but Microsoft, which uses OpenAI-developed models in its products, is a provider. As we discuss below, current AI watermarking techniques require the cooperation of model providers.
} For example, Google launched SynthID, a watermark embedded across its commercial AI products \cite{deepmind2025synthid}. The European Union's AI Act's Article 50, which goes into effect in August 2026, sets certain ``transparency obligations'' for both AI providers and deployers to provide infrastructure for labeling synthetic content that has not undergone human review. The exact technical form of this requirement has not yet been specified, but the Act states that providers must ``ensure that the output of the AI system are marked in a machine-readable format and detected as artificially generated or manipulated.'' \cite{EU_regulation_2026} Though the Act leaves several broad exceptions to the labeling requirement (for example, in the case of human review of the synthetic content), model providers will likely need to offer technical means for labeling outputs in the form of watermarks or other disclosure mechanisms. This large-scale regulatory push to develop and deploy machine-readable marks makes it all the more pressing to address the kinds of issues, which this paper raises, around the use of digital watermarks for AI governance.

In the US, the 2023 Biden Administration Executive Order on Safe, Secure, and Trustworthy Development and Use of AI specifically instructed government agencies to develop standards for labeling synthetic content by means of watermarking \cite{eo2023safe}. The order has since been revoked by the Trump Administration, which reversed course on the urgency of labeling \cite{eo2025removing}, but some state-level initiatives continued to build on the previous administration's efforts \cite{ca2024sb942, livingstone2025provenance}. These shifting regulatory trends suggest that rather than a broad solution for identifying and labeling synthetic content, we are likely to see the continued emergence of a patchwork of public and private orderings. But even in such a heterogeneous environment, watermarks are still often leaned on as silver-bullet solutions to the challenges of synthetic content.

Traditional watermarks are used to determine the authenticity of content, linking its provenance to its authority \cite{cox_first_2002}. But a watermark on synthetic content is used to determine \textit{in}authenticity. Operationally, this places slightly different requirements on an AI watermarking method: a traditional watermark should be difficult to forge, whereas an AI watermark should be difficult to remove. In this sense, AI watermarks are functionally opposite to digital signatures: a digital signature should break if a signed document is altered, whereas an AI watermark should be detectable even after some significant alterations.\footnote{ Other disclosure methods that rely on digital signatures of provenance statements are therefore distinct from digital watermarking. Though even the legal and epistemic status of cryptographically signed digital signatures was not always taken for granted as trustworthy. For more on the history of digital signatures in French and US courts, see \citet{BlanchetteJean-François2012Bop:}. We discuss the distinction between watermarking and cryptographic signature-based provenance tools in greater detail below.} Much of the critique of AI watermarking has focused on this inversion of the original use of watermarks as signs of authenticity. The digital media environment makes synthetic content malleable in the hands of the user, raising questions over their brittleness \cite{zhang2024watermarks}. Additionally, if only some AI model providers adopt watermarking schemes, what about media produced using self-hosted models or providers who opt out of a watermarking scheme? And even if a piece of media does bear an AI watermark, what sense should users and platforms make of the trustworthiness of this object \cite{leibowicz2023watermarking}? In short, critics argue that watermarks are technically brittle, epistemically ambiguous, and politically ineffective in addressing the challenge of trust in an information environment saturated with synthetic content.

These critiques, we argue, adopt a forensic perspective on the challenges of synthetic content. They primarily point to the shortcomings of watermarks as a means to authenticate the status of a given individual piece of media. Computer forensics (and the subfield of digital media forensics) belong to what literary and media scholar Matthew Kirschenbaum described as the ``forensic imagination'' of modernity---the desire to uncover meaning, identity and intentionality in the smallest traces, fragments, and overlooked details \cite{KirschenbaumMatthewG2008M:nm}. The technical and epistemic challenges of enacting the forensic imagination, however, predate the rapid spread of synthetic content and pertain to a long history of digitally or otherwise manipulated media \cite{chesney_deep_2019, Habgood-Coote2023-HABDAT-2, ScheirerWalter2024Ahof}. The use of watermarks as a forensic tool is indeed limited. But a particular challenge raised by the widespread availability of commercial AI tools is one of scale. ``The inversion'' as a tipping point for trust in digital media as a whole turns a question of assuring authenticity of a given media object into a question of measuring and contending with saturation.

Rather than the forensic hope to accurately and comprehensively signal the presence of individual pieces of synthetic content, we argue that digital watermarking is better suited to an ecosystems approach: their primary utility is in measuring system-level impacts of synthetic content as a whole on media environments. Much like epidemiologists use wastewater testing to track the spread of disease in particular geographic communities---instead of, or as a complement to, testing individuals---digital watermarking can provide a high-level view of synthetic content’s incursion into our media ecosystems.

Shifting to the ecosystems approach still raises challenging, but somewhat different, questions surrounding the deployment of watermarks and their uses in AI governance. We first provide a technical overview of audio, visual, image, and text-based watermarks. We then discuss five high-level governance questions that need to be addressed by policymakers, model developers, platform operators, and users if the ecosystem approach is to be used instead of, or as a complement to, the forensic approach. We argue that the ecosystem model offers more workable paths to resolving these governance questions than the forensic mode. Lastly, we ground the payoff of reconceptualizing AI watermarks as ecosystem indicators by examining two scenarios for their use in two types of media platforms---a music streaming service concerned with audio and a scientific publishing platform concerned with text-based generation of preprint publications.

\section{Technical Background}

Identifying synthetic content is challenging because genAI systems are engineered to mimic human content. AI systems ingest text, images, videos, and other media created by human authors---``training data''---and developers train the systems to generate content that is statistically indistinguishable from the training data. Popular AI detection methods attempt to identify synthetic content by noticing imperfections in the process used to train the AI \cite{gptzero, mitchell2023detectgpt, hans2024spotting}. These methods place AI detectors in an adversarial relationship with AI providers. Watermarking, by contrast, is not inherently adversarial. AI systems are not deliberately designed to deceive, but rather to generate high-quality content that is, by definition, difficult to distinguish from human content. AI watermarking methods formalize a collaborative relationship between providers and detectors: providers inject a subtle signal into synthetic content, the watermark, that detectors can use to confidently identify the synthetic content.

An AI watermark has three primary desiderata \cite{kuditipudi2024robust}. First, the watermark should make synthetic content easily detectable. Ideally, watermarking methods aim for strong statistical guarantees on the rate of false positives of AI detection---while it is difficult to provide guarantees for adversarial detectors, such guarantees are often possible for watermark detectors because the detector relies on statistical patterns intentionally embedded into synthetic content during generation, rather than heuristic markers of synthetic content that could coincidentally appear in human content. Second, the watermark should not degrade the quality of synthetic content. Neither the model provider nor the users of genAI will tolerate a protocol that materially diminishes the AI product (i.e., its ability to produce high-quality content). Third, there should be some amount of friction to remove a watermark from synthetic content. This final property (known as ``robustness'') allows for an adversarial relationship between the AI provider and AI users or distributors: while a responsible AI provider may implement a watermark to clearly identify synthetic content originating from their system, users of this system and content distributors may try to remove evidence of the content’s synthetic origin. Robustness precludes the use of metadata for tracking content provenance and entails that the watermark must be embedded directly into the content itself, as metadata can be easily altered, or separated from content entirely.\footnote{ For a survey of techniques for visual watermarking see \citet{begum2020digital} and for work on watermarking language models see \citet{kirchenbauer2023watermark}.}

These three desiderata are technically achievable because AI watermarks are inserted into synthetic content at its moment of inception, unlike traditional watermarking methods, which add watermarks to pre-existing content post hoc. By weaving the watermark into synthetic content as it is generated, it is possible to embed a strong statistical signal while guaranteeing the quality of generated content. Traditional post hoc methods seek to hide a watermark in imperceptible bits of the content, but these steganographic techniques are difficult to apply to media like text, where every bit is perceptible. Post hoc watermarks that rely on, for example, synonym substitutions make subtle alterations to the meaning of text, potentially degrading its quality.\footnote{For a review of post-hoc watermarks for text, see \citet{kamaruddin2018review}.} In contrast, recent work on genAI watermarks for text guarantees that text generated with a watermark has comparable quality to text generated without the watermark. In this setting, it is possible to make strong guarantees on quality---meaning text generated with a watermark is provably indistinguishable from text generated without a watermark, unless you have knowledge of the watermark and know exactly what to look for. \cite{christ2024undetectable,kuditipudi2024robust}

Crucially, detection of an AI watermark is a statistical endeavor: watermark detectors scan content for evidence of the watermark and return an assessment in the form of p-values that measure the probability of observing the content under the null hypothesis that the content was generated without a watermark.\footnote{In contrast to wastewater measurements, which must be calibrated against an independent source of truth, watermarking protocols control both insertion and detection of the watermark, meaning that detection is largely calibrated by construction. For empirical analysis of watermark calibration and robustness in-the-wild, see \citet{sander2026textseal}.} The smaller the p-value, the stronger the evidence of a watermark. The strength of an AI watermark is constrained by the entropy of the AI generator---high-entropy content like audio, images, and video is relatively easy to watermark compared to low-entropy content like text. This principle extends to more narrowly defined generation tasks. For example, while both computer code and creative writing can be produced by the same text generation system, computer code is typically lower-entropy than stories. Therefore, applying an AI watermark to a text generation system will typically impart weaker watermarks on generated code than it does on generated stories. Designing for weak signals affords some flexibility. For one, a stronger watermark may require adding so much information to the synthetic output (especially in the context of text generation) that outputs would be significantly distorted by the watermark. Trading off the strength of the watermark can minimize the watermark’s impact on the quality of generated content.

As a statistical measure, watermarks are distinct from other provenance tools that rely on a cryptographic signature appended to the piece of media's metadata. One prominent initiative to create infrastructure for digital provenance is the Coalition for Content Provenance and Authentication (C2PA) standard for ``Content Credentials.'' Backed by a coalition of news and media organizations, genAI tools developers, social media platforms, and academic publishers,~\footnote{\url{https://c2pa.org/membership/}} the C2PA Content Credentials standard offers a method of binding contextual information about the methods of production of a piece of media to its metadata. The cryptographic authentication the standard provides is not statistical but rather able to provide assurance that the information contained in the signed metadata is accurate. It is often presented as a complementary means of authenticating and providing information about the provenance of a piece of media to tools like watermarking (though some have argued the use of both tools may result in contradictions  \cite{nemecek2026authenticatedcontradictionsdesynchronizedprovenance}. A fundamental issue with the C2PA approach is that the metadata itself, though cryptographically signed, can be easily removed by a user before posting a piece of media to a platform checking for signed metadata that would label it as synthetic content. To effectively deter such acts of removal, an authenticator would need to treat the absence of such metadata as itself a reason to suspect content, meaning that provenance tools need to be applied to any piece of content, whether synthetic or not. This assumes a fairly stable and centralized content environment where parties are not adversarial and where the infrastructure for preserving and interpreting the metadata is widely available across a variety of content hosts. For these reasons, rather than focus on how cryptographic signatures may provide greater degrees of forensic certainty, we focus on the ways watermarks may be more effective measures of ecosystem-level questions.

\section{Reconceptualizing Digital Watermarks}

As synthetic content flows into different media environments---social media, news, streaming music, film, academic scholarship---it is creating different problems for different participants in each space. On social media, for example, users and content moderators have to judge the authenticity and veracity of each text, image, and video they encounter, while the platforms organizing and distributing that content have to judge whether it violates their terms of service or community guidelines, and identify bots and ``coordinated inauthentic behavior'' by malign actors. On music streaming platforms, synthetic content---AI-generated music---threatens to capture a share of listeners’ attention, hurting artists’ ability to reach and profit from new audiences. And if the AI-generated music is bad, it threatens to turn users off of platforms, or to reduce the amount of time they spend on them. In the film industry, writers, actors, musicians, and artists are contending with synthetic content that appropriates their image, voice, style, ideas, and other creative contributions. In academia, peer reviewers and publishers struggle to enforce norms of academic integrity as synthetic content is incorporated into the research process.

Given the myriad ways synthetic content can impact individual and collective rights and interests, it’s unlikely that any single, comprehensive approach to governing such content will emerge. Rather, in the coming years, we can expect to see a heterogeneous mix of public policy and private ordering develop in piecemeal fashion to resolve specific kinds of conflicts between specific kinds of actors. Policymakers will deploy what legal tools they can in response to pressure from constituents and industry groups, while courts decide how existing laws---in areas such as intellectual property and product liability---apply to AI-related cases. Companies will develop technological solutions, devise new licensing and contractual arrangements, and invest in brand management to avoid alienating users. Unions and civil society groups will organize, protest, audit, litigate, educate the public, and engage in other forms of collective action, while individuals test whatever self-help strategies they can muster. 

The question for proponents of digital watermarks is where they fit---what role, if any, they can play---in this rapidly changing, unchoreographed ensemble. Watermark enthusiasts have traditionally answered that they are forensic tools for reliably identifying individual pieces of synthetic content. We argue that critics of digital watermarking are right to view this standard answer with skepticism. But that does not mean digital watermarking has no role to play. Instead of a forensic conceptualization of AI watermarks, we argue for an \emph{ecosystems perspective}---an approach that uses the centralized nature of AI model providers and the statistical nature of digital watermarking as indicators of overall synthetic content saturation in a given information ecosystem. A watermark applied en masse to synthetic content can be used to assess the population-level impacts of content originating from an AI provider on that ecosystem, even if it is only weakly detectable. Recognizing the usefulness of a brittle watermark is especially significant for considering the usefulness of watermarks for textual synthetic content, where weaker signals are expected. Rather than asking watermarks to answer the forensic question, ``is this AI?,'' we propose to use them as indicators of how much AI is all around us.

In what follows, we show why researchers, advocates, and policymakers should reconceptualize digital watermarks as tools for understanding media ecosystems rather than forensically identifying specific pieces of synthetic content. And we explore how the challenges associated with using watermarks to govern AI are transformed---and made more tractable---when the utility of watermarks is reconceptualized in this way.

\subsection{Watermarks as Evidence}

Watermarks are signals---evidence---about the provenance of the media they’re attached to, which have to be interpreted and acted on by some audience. To use watermarks effectively as a tool for AI governance, the first questions to ask are what kind of evidence they provide and to whom they provide it.

As we discussed above, the evidence digital watermarks generate is statistical---it indicates the probability that content is synthetic\footnote{More specifically, as we discuss above, they indicate the likelihood, expressed as a p-value, of detecting a watermark signal as strong as the one detected in an individual piece of content known to not be synthetic.
}---and intrinsic to statistical evidence is the possibility of error. Operating in the forensic mode, someone using digital watermarks to discern if a particular piece of content was AI-generated must therefore contend with the possibility of false positives (wrongly concluding that real content is synthetic) and false negatives (wrongly concluding that synthetic content is real). For trained statisticians this is not a problem, as there are methods by which to systematically account for statistical errors. But for the lay interpreter of digital watermarks there is significant room for misunderstanding. The average person is notoriously bad at reasoning about probabilities. If a college professor is told, for example, that there is a 68\% likelihood a student’s essay contains synthetic content, there is a serious chance the professor will over-interpret the signal and conclude with too much confidence that the student cheated (potentially a false positive). Similarly, if the statistical threshold for flagging posts on social media as ``possibly AI-generated'' is set too high, casual readers might assume that un-flagged content is necessarily real (potential false negatives).

For digital watermarks to be used responsibly in a forensic mode---i.e., to identify individual pieces of synthetic content---care must be taken to ensure that they are contextualized appropriately and interpreted correctly. For example, courts routinely consider statistical evidence and there are long-standing debates among law scholars about how to properly incorporate such evidence into legal proceedings. One can imagine effectively introducing forensic digital watermarks as evidence at trial (to support claims about, say, intellectual property rights or legal liability in relation to AI-generated content), where attorneys are required to explain and contextualize the watermark, experts can be called to contest their interpretation, and disinterested judges are responsible for overseeing the process of determining their meaning.\footnote{Science and Technology Studies scholars have developed a robust critique of the authority of forensic evidence in courts \cite{ColeSimonA.2001Si:a,RobertsDorothy2011Fi:h} and the commodification and advocacy role of expert witnesses in interpreting scientific evidence \cite{GoodwinCharles1994PV,JasanoffSheila1995Satb}. So while we use here the court as illustrative of a site where institutional scaffolding for evidence interpretation is provided, it too can serve as a site where evidence is over indexed and misinterpreted.} In many other contexts, however, including those---like plagiarism detection---where enthusiasm for these tools is most pronounced, there is less institutional infrastructure to guide their implementation.

Rather than confine their use to the handful of institutional settings equipped to interpret them correctly, however, we could put digital watermarks to other uses. We could use them to answer a different question than the forensic approach asks, one that’s a better fit for the kind of statistical evidence they offer. Instead of asking ``is this piece of content real or synthetic?'' we could ask ``how is synthetic content impacting this media ecosystem, on the whole?'' Using watermarks to assess the provenance of individual pieces of content requires a nuanced understanding of false-positive rates, hypothesis testing, and uncertainty quantification. Using watermarks to study content flowing through  informational ecosystems in the aggregate avoids much of this nuance. To answer questions about the aggregate---taking what we call an ``ecosystems approach''---statistical evidence is exactly what you need. 

Having recast the questions digital watermarks aim to answer, different governance interventions come into view. Forensic questions, such as whether a given piece of content is synthetic, which model generated it, and who is responsible for its creation, are conducive to individualized interventions (holding specific parties accountable for harmful content, for example, or asserting intellectual property rights). By contrast, ecosystem questions lend themselves to systems-level interventions, addressing broader patterns and structures of information environments (such as bringing down aggregate levels of synthetic content). Consider, again, the case of plagiarism detection. Forensic questions---e.g., ``Was this paper generated by AI?''---suggest a particular kind of response: if it was, then charge the student who submitted it with an academic integrity violation. Ecosystems questions---e.g., ``How much student work at our university is AI-generated?''---point in a different direction. If the problem is overall levels of plagiarism and the goal is to reduce them, then one might intervene by rethinking how to evaluate student performance or by reconfiguring the incentives students face.

\subsection{Watermarks as Friction}

Shifting from a forensic perspective to an ecosystem perspective can also help address concerns about the robustness of digital watermarks. Skeptics warn that watermarks are brittle---too easily undone by anyone determined enough to circumvent them \cite{sadasivan2023ai-generated, zhang2024watermarks}. But if the goal is not simply to identify and address each individual piece of synthetic content, and instead, to understand and intervene at the ecosystems level, then worries about robustness lose force. Although any individual watermark might be vulnerable to tampering, removal, or avoidance, it would be complicated, even for skilled attackers, to disable watermarks at scale. 

Indeed, viewed from this vantage point, questions about robustness start to resemble more familiar issues around information and internet governance. In debates about how to manage online mis- and disinformation, for example, scholars have argued that, for both practical and legal reasons, the goal of perfect enforcement---eliminating all misleading content---is out of reach. Instead, they suggest, policy interventions should aim to lower overall levels of mis- and disinformation, either by increasing the costs of producing and disseminating it, or by slowing the process of ingesting it (and thereby reducing the odds it’s believed). Both goals can be advanced, they argue, by introducing friction into online information environments \cite{goodman2021digital}. 

Traditionally, human-computer interaction research aimed to minimize friction. The best user interfaces were ``seamless''---they allowed users to work without interruption, to become absorbed in their activity as the computer became, in design theorist Don Norman’s language, ``invisible'' \cite{norman1999invisible}. Friction, according to this view, was the enemy. But as the risks of engaging uncritically with computers (especially online) became more evident, the idea of encouraging people to pause, giving users time to think and reflect, gained purchase. HCI scholars have begun to advocate for friction as a mechanism to create ``mindful interactions''---``interactions that are reflective, informed and safe'' \cite{cox2016design}---and legal scholars have begun to theorize various forms of ``desirable inefficiency'' as a means of ``inject[ing] complex human values into systems'' \cite{ohm2018desirable, frischmann2023friction}.

Critics of watermarks are right to insist that they lack robustness---that they can't guarantee perfect enforcement. But watermarks need not be perfectly robust to create friction in content production, circulation, and consumption \cite{fernandezWhatLiesAhead2024}. On the supply side, watermarks can raise the costs of generating and distributing synthetic content, perhaps enough to disincentivize at least some socially undesirable uses, and they can help slow the spread of harmful content. On the demand side, watermarks could provide evidence of the overall prevalence of synthetic content, providing consumers of content with a general barometer of how to approach content on the platform with more careful consideration, caution, or skepticism. In the case of social media, for example, rather than expect watermarks to provide a perfect infrastructure for labeling synthetic content (a challenge in both accuracy and interpretation, as we discuss further in the next section), providing users with a public-facing measure of the overall prevalence of synthetic content on the platform might be a mechanism for overriding the seamlessness of the interface.

The desire for perfect enforcement is often also accompanied by calls for automated enforcement---the scale of online content requiring some clear means of triggering content removal or posting without much human judgement in each individual case. The forensic hope for watermarks in such a regulation-by-machine scenario amounts to an ex ante measure---removing watermarked content before it gets posted or labeling it at the moment of posting. Early experiences of automated enforcement online, especially the use of digital rights management (DRM) tools to enforce intellectual property rights, are a cautionary tale for ex ante approaches, leading many legal scholars and critics to prefer ex post strategies for regulating digital media, which allow for flexibility and interpretive openness of legal doctrines such as Fair Use \cite{ZittrainJonathan2008TFot,44ae0609-7b86-3ed5-bf5d-b48a2c878dc6}. As Tarleton Gillespie put it, ``instead of using law's ex post assessment of intent and consequence, technology prepares for the worst and lives by it''~\cite{GillespieTarleton2007Ws:c}. The ecosystem approach we advocate is more compatible with ex post assessments of the meaning and significance of watermarks, compared with the forensic hope for perfect, ex ante enforcement. As discussed in the ``Watermarks as Evidence'' section, above, appropriate institutional scaffolding is a precondition for interpreting the meaning and strength of claims made based on the statistical result a watermark detector yields. But the kinds of ex post strategies most likely to succeed are not geared toward flagging individual pieces of content as synthetic; rather they aim to offer insight into the overall levels of synthetic content in a given information ecosystem.

\subsection{Who are Watermarks For?}

The first requirement for watermarking synthetic content at scale is for model providers to embed watermarks in their products' outputs. But it is not immediately obvious who ought to be the audience for that signal and through what detection mechanism. For example, Google's SynthID system does not share widely the technical specs of the SynthID watermark, but it does offer a publicly available detector anyone can use to pick up the watermark’s signal. On the other hand, Meta, as a content platform, uses both its own and other providers' watermark and metadata standards to label some synthetic content as AI-generated across its platform, but does not offer a public-facing detector like SynthID. More recently, OpenAI announced the launch of its own public detection tool which relies on C2PA, SynthID, and other undisclosed watermarking technology \cite{OpenAi2026}.

There are good reasons to keep a watermark system and its detector closed. Knowledge of the specific technical mechanism is one way in which a user wishing to remove the watermark can circumvent detection. But even a blackboxed detector like Google's can allow an adversary to query it enough times to derive some insight into possible effective circumvention. This dynamic is what Leibowicz et al. refer to as the ``detection dilemma'': the more widely available detection tools become the less effective they are, as more is learned about how to circumvent them \cite{leibowicz2021deepfake}. But the dilemma is primarily acute in the forensic mode.\footnote{The authors lay out seven distinct detection contexts across different scales of media and differently resourced detecting actors. But in all these contexts, the question still remains one of specific media authenticity as part of a forensic question, and not one of saturation of the overall ecosystem or even a given platform \cite{leibowicz2021deepfake}.} As discussed in the ``Watermarks as Friction'' section, the labor-intensive process of deriving possible strategies for circumvention from the availability of a detector is less of a problem when we want to measure both weak signals and overall prevalence at scale. The problem of access then is not about the detection tool itself, but about access to a sufficiently representative sample of content in the overall ecosystem.

Examining proprietary technological systems from the outside has long been a challenge in designing procedures for auditing and evaluating both AI tools and digital media platforms. The individual experience of the researcher-as-user is an unstable representation of the system as a whole given the role played by customization, algorithmic curation, and the indeterminate nature of a model's output \cite{SeaverNick2019KA,burrell2015opacity}. Even with aggregate evaluations among larger test groups, the experiences of certain underrepresented groups within the test group may go unnoticed \cite{barocas2021designing}. An ecosystems approach might therefore be most useful in contexts where the content host or ``owner'' of the ecosystem is motivated to measure the pervasiveness of synthetic content. One clear example where these interests align is the academic plagiarism case discussed above, where a university is interested in a bird's-eye view of work submitted by students. As we explore in one of the scenarios below, streaming platforms or social media platforms may also have reasons for wanting such insight and producing an internal detection process. And though internal audits are fraught tools for creating external, public accountability, as Raji et al. argue, they can help by producing shareable artifacts like transparency reports \cite{raji2020closing}.

Turning away from the forensic perspective also reframes the question of effectively communicating the meaning of watermarks to audiences directly consuming watermarked content. When social media platforms sought to apply synthetic content labels to user-shared content, they quickly discovered that user expectations differed from the taxonomy of 
labels. For example, Meta reversed course on its initial labeling policy after users complained that authentic images that were then retouched or edited using AI tools were being labeled as synthetic content as though they were wholly generated by an AI tool \cite{bickert2024approach}. Ecosystem approaches raise different questions about how users of a platform might interpret the platform-level disclosure of the rates of synthetic content. It may cultivate a more skeptical or more trusting orientation towards any given piece of media they encounter if the rate is either higher or lower than their expectations. Disclosing that synthetic content is being monitored may also lead to greater trust in the platform's proactive governance measures or suspicion toward its monitoring of user behavior. It may also be taken up as a competitive advantage that draws users to certain platforms that provide such an analysis. In none of these cases do watermarks ``solve'' the problem of trust in the platform. But unlike the forensic approach, the ecosystems approach draws user attention to their overall media environment, rather than providing an ambiguous and potentially misapplied label to specific pieces of content.

\subsection{Risks of Watermarks}

Using watermarks as a tool for governing AI is not without risks. Thus far, we have focused on the risks that stem from using watermarks in the standard forensic mode---especially, as we discussed in the ``Watermarks as Evidence'' section, the risk of misinterpreting the statistical evidence watermarks provide, which is likely in contexts where appropriate interpretive infrastructure is lacking. But using watermarks in the ecosystem mode that we argue for carries risks too, both at the individual level and at the population level.

At the individual level there are risks to privacy, particularly for high-bandwidth information channels like video content. For high-bandwidth channels, not only can a watermark be detected in individual units of synthetic content, watermarks could be personalized to an individual user of a genAI system \cite{cohen2024watermarking}. A personalized watermark identifies not just the originating model for a unit of synthetic content, but also the specific account with the model provider used to generate the content. Obviously, any tool that can be used to identify content generated by a particular person or group can be used to surveil them. Using personalized watermarks in this way would fit a recurring pattern---exemplified, most notably, by the ``surveillance advertising'' model---of turning otherwise innocuous digital traces into tracking devices. Indeed, it’s possible that their value for surveillance could ultimately be the economic incentive that drives widespread adoption and deployment of genAI watermarks. Importantly, though, personalized watermarks are much less realistic for low-bandwidth channels like text, where there is simply not enough entropy in the generative process to imbue content with a strong personalized watermark, compared with watermarks for images or video.

At the population level, digital watermarks risk creating complacency. Because they are easy to implement and (by design) have minimal impact on the quality of a genAI system, watermarks could be seen as a quick fix---a means for policymakers to demonstrate responsiveness as synthetic content floods our information ecosystems, in a manner that comes with very little cost. While watermarks can improve our ability to monitor flows of synthetic content through these ecosystems, they do not address more difficult policy questions about how to actually regulate those flows. Furthermore, watermarks require the compliance of genAI providers to deliver watermarked content. Even in a world where major model providers agree to watermark their offerings, there will be blind spots. This includes unwatermarked synthetic content from bespoke model providers, as well as synthetic content that has been scrubbed of watermarks by adversarial actors \cite{zhang2024watermarks}. Though adversarial scrubbing is more likely in scenarios where watermarks are deployed for forensics or moderation (banning content that meets some threshold of watermark detection) than in contexts where they are used for monitoring and ecosystem analysis.

At both the individual and population levels, there is the additional risk that watermarks could be forged. Most current AI watermarking protocols depend upon a watermark key that is used by both the generator, to synthesize watermarked content, and the detector, to identify the watermark in synthesized content. Any actor in possession of the detector has access to this key and can therefore synthesize new watermarked content with the same watermark as the original model provider. We can imagine any number of incentives to forge a watermark---one salient analog is the forgery of User-Agent strings by web browsers. Methods that isolate the detection capability from watermark generation are an active area of research \cite{liu2024unforgeable,fairoze2025publicly} but their practicality is uncertain and theoretical results suggest that they are unlikely to be very robust \cite{fairoze2025difficulty}.

Lastly, watermarks may be developed using skewed human-generated content that could lead to a bias in false positives (human-generated content being flagged as synthetic content). In text-based watermarking in particular, there is a concern that the charge of ``sounding like AI'' might be directed at particular groups such as English language learners or minoritized groups. Prior work on bias in AI systems such as automated speech recognition \cite{doi:10.1073/pnas.1915768117} highlights this potential for disparate outcomes in analyzing human speech and text generation. If watermarks are used in the ecosystem mode, studies about the potential systemic bias in false positives will be a necessary step in assessing the system's fairness.

\subsection{Incentives For and Against Watermarks}

Finally, whether digital watermarks are a viable governance tool will be determined, in part, by how relevant stakeholders understand incentives for and against adopting them. To conclude this part of the paper, we thus turn to how incentive structures for watermarking change when watermarks are used for ecosystem monitoring rather than forensic identification. By exploring some common, if stylized, arrangements we can discern patterns of incentives where the ecosystem mode may lead to a stronger alignment of interests for watermarking synthetic content. The forensic mode often assumes an adversarial relationship between the content producer and content host or viewer. We argue that using watermarks in the ecosystem mode, rather than the forensic mode, could reduce the incentives of some parties to resist their adoption and ``bring down the temperature'' in situations where incentives for and against watermarks remain misaligned, thereby increasing the odds they’re implemented.

First, consider ``enterprise'' or ``business-to-business'' (B2B) settings. These are cases where AI firms sell AI services to other companies. For example, companies buy access to chatbots to automate customer service work and they buy access to AI coding tools to automate or partially automate in-house software development. Such B2B arrangements are increasingly the dominant place where frontier AI companies are looking to monetize their products and recoup the astronomical investments being made in model development  \cite{isaac2026samaltman}. 

If watermarks are understood as forensic tools, the incentives landscape in these cases is likely to be mixed. Companies buying AI services are likely to welcome watermarks, since they have an interest in gauging how much the services they’re paying for are utilized and how much value they’re generating for the company. AI providers might have competing motivations---eager to attract and satisfy customers, but hesitant to divulge evidence, where it exists, that their services are under-performing---but they are likely, in the end, to give their customers what they want.\footnote{ Indeed, it’s easy to imagine AI providers using watermarks as a means of differentiating their services from competitors. In the context of targeted advertising, for example, companies like Google and Meta have worked tirelessly to develop tracking technologies that provide their customers---advertisers---with evidence that their advertisements work. See \cite{mcguigan2023selling}.} By contrast, the company employees who are expected to use AI might resist, worrying that watermarks could be used as instruments of worker surveillance (as discussed in the previous section). But notice how the incentives change if watermarks are deployed for ecosystem monitoring rather than forensic identification: if individual employees aren’t at risk of surveillance through watermarks they will have less reason to oppose them.

Next, consider cases where the AI provider and the platform for generated content are one and the same: for example, social media companies like Meta and TikTok have deployed their own generative AI systems for use by their own users in their own apps \cite{meta2024llama3,tiktok2024symphony}. In such cases, both the platforms and a large segment of their users---those who primarily consume content---are likely to embrace watermarks. The platforms will want internal data about how their technology is being used, and content consumers will want to know when they’re viewing synthetic media. If watermarks are deployed as forensic tools, however, another segment of users---content producers---may attempt to disable or evade them in order to pass off synthetic content as their own. Notice again, though, how incentives change when watermarks are used for ecosystem monitoring: if platforms only aim to detect aggregate levels of synthetic content, and content creators don’t face the threat of identification and sanction, no individual creator will be motivated to undertake the considerable effort required to circumvent detection.

Lastly, consider cases where AI models are used by consumers to generate content distributed elsewhere: LLMs are used to write posts for social media; music generation tools are used to create songs distributed on streaming platforms; students use AI to complete homework assignments; or researchers use AI to write scientific papers. In these kinds of cases, incentives are likely to remain misaligned regardless of how watermarks are conceptualized and utilized, since content consumers will be interested in knowing what to make of what they’re seeing and hearing, while AI providers will have little motivation to generate evidence---either at the individual level or in the aggregate---that their tools are implicated. 

Even in these cases, deploying watermarks for ecosystem monitoring rather than forensic identification could help defuse tension and motivate powerful actors, such as rightsholders or regulators, to insist on their adoption. For example, Spotify---likely in response to pressure from the recording industry---announced new efforts to help ``[preserve] trust across the entire music ecosystem'' by better managing the flow of AI-generated songs.\footnote{\url{https://newsroom.spotify.com/2025-09-25/spotify-strengthens-ai-protections/}} More broadly, lawmakers often look to mandated disclosure as a low-cost and light-touch means of regulation \cite{susser2019notice}. While powerful AI companies are sure to resist policies compelling them to label individual pieces of content as having been generated by their own systems, they may be less inclined to spend political capital fighting against forced disclosure of aggregate, ecosystem-level statistics.

\section{Scenarios}

To understand the implications of ecosystem analysis using watermarks, we consider the examples of (1) a music streaming service and (2) scientific research. While these scenarios are speculative, they provide more granularity and context for assessing implications of the ecosystem perspective. In the case of music streaming, content consists of high-bandwidth audio, so forensic analysis of fully synthetic, watermarked content may sometimes be possible; but as we discussed in Section 3.1, the mere existence of statistical evidence does not tell us how and when to act. An ecosystem analysis answers a distinct category of questions about the prevalence and distribution of synthetic content on the service. In the case of scientific research, content is low-bandwidth text (i.e., papers) making any forensic analysis more ambiguous. Here an ecosystem analysis can provide critical insight into prevailing patterns of synthetic content.

\subsection{Music Streaming Service}

A music streaming service has three sets of stakeholders: the users of the service, artists who distribute their music via the service, and the service provider itself. The interests of these stakeholders may be misaligned with respect to synthetic content, raising the questions of competing interests discussed in Section 3.3. Early responses from users express a strong preference for authenticity in the initial wave of synthetic content \cite{getty2024trust,parshakov2025users,kirk2025aauthorship}. Artists see synthetic content as competition that devalues their work \cite{goetze2024ai}; synthetic content has expressly been used to dilute human artist revenue in scams that take advantage of the monetization strategy of streaming services \cite{us_v_smith}. The service provider itself may benefit from synthetic content if it reduces licensing costs in the short term, though the ubiquity of synthetic content on its platform may undermine its reputation and impact its subscription model in the long term. This echoes the tensions in incentive structures discussed in Section 3.5.

We expect that both forensic and ecosystem analyses will be of interest to the stakeholders of music streaming services. Audio is a medium-bandwidth information channel—higher bandwidth than text, but lower bandwidth than video. Forensic identification of watermarked, fully synthesized audio recordings might sometimes be possible; forensic analysis of audio produced or edited with AI assistance will produce more ambiguous results. Editing operations are low entropy operations compared to full audio synthesis, and a watermark imparted on edited audio will be relatively weak. Weak watermarks are poor evidence for judicial decisions like content removal or demonetization, but these signals will still be statistically evident in the aggregate. If a platform were to publicly report on synthetic content within music categories or monetization pools, this disclosure could influence the behavior of users, advertisers, and distributors even without more heavy-handed forensic enforcement; this is the friction function of watermarks described in Section 3.2.

An ecosystem analysis would provide broad insight into the volume of synthetic content on a music streaming service. But producing such insight requires the cooperation of the platform to review content across its services, which may not align with the platform's incentives. This illustrates the access questions raised in Section 3.3; ecosystem analysis requires not just a detector, but also a representative sample of content from the platform. Still, the platform itself may have reasons to want such an analysis, revealing the relative volume of synthetic content across different music genres, as well as the relative consumption of synthetic content among different sub-populations of users. This transforms watermarks from a forensic tool for platform moderation into an analytical tool for platform governance, informing policy decisions about recommendation algorithms, playlist curation, and monetization strategies, as well as disclosures to other stakeholders discussed in Section 3.4.

\subsection{Scientific Writing}

The growth of synthetic content in scientific writing poses challenges to the integrity of peer review and moderation processes. For example, arXiv, a pre-print platform for scientific papers, no longer accepts preprints of computer science surveys and position papers due to an overwhelming rise in AI-written submissions \cite{boboris2025arxiv}. Using the ecosystem mode of analysis described in Section 3.1, arXiv determined that the moderation burden posed by an onslaught of low quality survey articles outweighs the value of hosting pre-review content in these categories. This determination was made based on heuristic assessment of the growth of synthetic content in these categories; watermarks could provide more nuanced and quantitative support for such decision-making.

Ethically, reasons can be marshalled for and against banning or marking the use of AI in scientific writing \cite{abernethy2024parrots}. Pre-generative AI tools like spellcheck and grammar suggestions have been a productive and valuable component of scientific writing for decades. There is no bright line between the use of generative AI as a writing assistant and its use to synthesize more substantive arguments and ideas,\footnote{See \cite{csiszar2024blurry} for a historical perspective on search aids and authorship in scientific writing.} though some venues have taken a strong position against the use of LLMs in academic writing.\footnote{See, for example, the editorial policy for the journal Ethics: \url{https://www.journals.uchicago.edu/journals/et/ai-policy}} Pragmatically, as arXiv's experience proves, an unchecked permissive stance toward frictionless production of synthetic papers threatens to overwhelm the peer review and moderation functions of scientific institutions \cite{doi:10.1126/science.adw3000}. While there are nascent efforts to assist peer review and moderation itself with AI,\footnote{See \url{https://agents4science.stanford.edu/} for a maximalist perspective on the use of AI for both scientific writing and peer review.} ultimately the integrity of the scientific process and trust in scientific institutions rests on human oversight.\footnote{Historians of science, however, suggest that the practice of peer review itself is a rather recent development designed more toward assuaging broader public critiques of the integrity of the institutions of science than for providing scrutiny of any specific article, echoing our proposal for the usefulness of the ecosystems approach \cite{csiszar2016peer}.} Reintroducing some friction into the authoring system in the form of watermarks, as discussed in Section 3.2, can support the integrity of these processes.

ArXiv has recently begun using hallucinated references as a natural watermark on AI-generated documents \cite{chawla2026arxivban}. While useful as a stop-gap, this is an ephemeral watermark that will disappear as AI systems continue to improve. More importantly, arXiv has taken an entirely forensic view of this natural watermark, using hallucinated citations as forensic evidence to support banning authors from the arXiv platform. The ensuing controversy illustrates the tensions in the forensic mode discussed in Section 3.5.

An ecosystem analysis of AI use in scientific writing, facilitated by watermarks, could assist with institutional oversight while avoiding the fraught binary decisions associated with a forensic analysis of whether a particular scientific document has been overly authored by AI. This is the change in perspective described in Section 3.1, from asking whether a paper is AI-generated to asking how AI use is changing the publication ecosystem. Ecosystem analyses could help guide policy surrounding the use of AI, for example, providing guidelines for authors and reviewers regarding the use and misuse of AI and assessing compliance with these guidelines. These analyses could also provide transparency surrounding the use of AI in scientific writing, assuring the integrity of the scientific process both for the broader public and for the scientific community itself.

\section{Conclusion}

This paper has advocated for ecosystem analyses of the distribution of synthetic content on digital platforms using watermarks. Skepticism towards the efficacy of watermarks has primarily been animated by the forensic imagination of synthetic content identification. While we agree that watermarks have significant limitations as a forensic tool for determining the individual case of synthetic content, we argue that they are well-suited for measuring and monitoring the flows of synthetic content in an information ecosystem, such as a media streaming platform, a dating site, or an academic institution. Taking the ecosystems perspective on watermarks can provide us with aggregate insight about synthetic content saturation, with lower technical demands on the strength and precision of the watermarking methodology. This perspective can yield insights even for low-bandwidth information channels such as short-form text, where AI watermarks are inherently weak.

Reframing watermarks from a forensic tool to an ecosystem indicator points to new uses and efficacy but it also changes the nature of the governance questions posed by watermarking technology. Rather than content moderation and enforcement, this ecosystem perspective asks us to measure and act upon aggregate watermark signals at scale. Policymakers and platforms should consider how synthetic content saturation statistics gleaned from watermarks might be disclosed to users, regulators, and researchers. Such evidence is also more likely to inform institutional changes to policies and incentives, rather than individualized punitive measures. This approach also places responsibilities upon AI model providers; for watermarks to function as an ecosystem metric, they must be imprinted upon a critical mass of synthetic content. From this perspective, while concerted top-down regulatory efforts seem to have halted, the current centralization of the AI model provider market suggests coordination and buy-in might still be feasible; convincing just a few major providers to implement AI watermarking would result in watermarks on a significant fraction of all synthetic content. At the ecosystem level, these signals would be far more salient than trace amounts of unmarked synthetic content produced by self-hosted models.

As adoption of genAI tools continues to grow, synthetic content will become increasingly pervasive and the line between fully human generated content and synthetic content will blur. Marking and labeling synthetic content with sufficient gradation and nuance at scale is increasingly challenging. Furthermore, attitudes towards the legitimacy and desirability of synthetic content may also change over time. In this emerging media environment, our goals may shift from identifying and policing synthetic content, towards a broader understanding of how genAI technologies shape our platforms and our discourse. Watermarks offer us a tool for navigating and adapting to a new epistemic reality in which synthetic content is not deemed inherently untrustworthy or antithetical to content produced by humans. But, while we have argued that watermarks have a role to play in this emerging media landscape, the ecosystem perspective also makes it clear that they are no silver bullet to the thorny challenges of our information environment. The key to effectively using watermarks is to ask the right set of questions.

\bibliography{references}

\end{document}